\documentclass[12pt]{article}
\usepackage[dvips]{graphicx}
\usepackage{pdproc}

  \makeatletter 
  \def\@cite#1{[#1]} 
  \makeatother    
\begin{document}

\renewcommand{\thefootnote}{\alph{footnote}}

\begin{flushright}
{\small KAIST-TH 2004/14}
\end{flushright}

\title{Fayet-Iliopoulos terms in five-dimensional orbifold supergravity}

\author{Hiroyuki Abe}

\address{ 
Department of Physics, 
Korea Advanced Institute of Science and Technology \\
Daejeon 305-701, Korea 
\\ {\rm E-mail: abe@hep.kaist.ac.kr}}

\begin{center}
\footnotesize
Talk presented at the 12th International Conference on Supersymmetry and\\ 
Unification of Fundamental Interactions, June 17-23, 2004, Tsukuba, Japan 
\end{center}
\abstract{
We derive an off-shell formulation for the boundary 
Fayet-Iliopoulos (FI) terms in locally supersymmetric 
$U(1)$ gauge theory on 5D $S^1/Z_2$ orbifold. 
Some physical consequences of such FI-terms, 
e.g., the generation of 5D kink mass for hypermultiplet 
are studied within the full supergravity framework.}

\normalsize\baselineskip=15pt

\section{Introduction}
When we consider physics beyond the standard model (SM), 
the supersymmetry (SUSY) is one of the promising candidate 
which can stabilize the electroweak scale, and achieve better gauge 
coupling unification at the grand unification scale. 
One of the remarkable feature is that the SUSY breaking is 
parameterized by $F$ and $D$ order parameters which is the 
auxiliary component of the chiral and vector multiplet respectively. 
As is well known, for instance, the tadpoles of the $F$ and $D$ 
are SUSY invariant. These terms cause O'Raifeartaigh and 
Fayet-Iliopoulos (FI) mechanism of SUSY breaking, respectively. 

Another perspective beyond the SM is models with extra dimensions. 
We extend not only the internal space but also the space-time itself. 
In this direction, orbifold models are particularly interesting 
which makes a correlation between the internal and the external space. 
In such model because of the nontrivial parity in extra dimension, 
we naturally obtain the quasi-localization of fields that results in some 
hierarchical structures in the 4D effective theory. Furthermore they 
provide some SUSY breaking and the sequestering mechanism by the spatiality 
of extra dimension, and the various mediation mechanisms from the hidden 
to the visible sector. 

Here we will focus our attention on the Fayet-Iliopoulos term in 
orbifold models. Let us consider the 5D $S^1/Z_2$ orbifold model 
with global SUSY where the orbifold direction is chosen in the 
$\sigma_3$ direction of $SU(2)_R$. In this case, we have two 4D fixed 
plane with $N=1$ SUSY. Then it was pointed out in 
Ref.~\cite{Barbieri:2002ic,GrootNibbelink:2002wv} 
that the $U(1)$ vector multiplet has FI-term at the fixed plane 
in the form of $\Big( \xi_0 \delta(y)+ \xi_\pi \delta(y-\pi R)\Big) D$ 
where $D$ is the $N=1$ auxiliary field that consists of the gauge scalar 
$\phi$ and the third component of the $SU(2)$-triplet auxiliary field 
$Y^{(3)}$ in 5D $U(1)$ vector multiplet as $D=2Y^{(3)}-\partial_y \phi$. 
If the FI-coefficients $\xi_0$ and $\xi_\pi$ take the integrable form 
$\xi_\pi=-\xi_0 \equiv \xi$, the $D$-flat condition gives {\it the periodic 
sign-function} $\epsilon(y)$ type vacuum expectation value (VEV) of the gauge 
scalar field $\langle \phi \rangle = \xi \epsilon(y)$. This VEV produces the 
kink mass for the charged hypermultiplet and results in the quasi-localization 
of the hyperino field~\cite{Barbieri:2002ic,GrootNibbelink:2002wv,Marti:2002ar}. 
In this manner, FI-term is generically important in global orbifold models. 
Then our question is about the situation in local supersymmetry. 

In 4D supergravity (SUGRA), the FI-term is forbidden by the local 
SUSY except the case of gauged $U(1)_R$ or so called pseudo-anomalous 
$U(1)$. However in 5D orbifold SUGRA, the on-shell analysis in 
Ref.~\cite{Barbieri:2002ic} showed that the existence of some {\it bulk} 
$Z_2$-odd operator with the $\epsilon(y)$-type coefficient implies the 
existence of {\it boundary} FI-term for a (normal) $U(1)$ vector multiplet. 
But the exact off-shell formulation for such FI-term in 5D orbifold SUGRA 
was absent. In our work~\cite{Abe:2004yk}, we derive the off-shell formulation 
of such boundary FI-term by utilizing the off-shell 5D SUGRA~\cite{Fujita:2001bd} 
with four-form Lagrange multiplier~\cite{Bergshoeff:2000zn} which can generate 
the $\epsilon(y)$-type coefficient in a consistent way with the local SUSY.

\section{Formulation}
In the off-shell formulation of 5D SUGRA~\cite{Fujita:2001bd}, 
we introduce three $U(1)$ vector multiplets 
${\cal V}_A=\big(M^A,A^A_\mu,\Omega^{Ai},Y^{Aij} \big)$ where $A=Z,X,S$. 
We use the notation for the gauge scalar fields as $M^A=(\alpha,\beta,\gamma)$ 
with $Z_2$-parity $(+,-,-)$ respectively. The ${\cal V}_Z$ is the graviphoton 
multiplet, ${\cal V}_X$ is the multiplet which finally has boundary FI-term 
and ${\cal V}_S$ will be eliminated by the constraint generated by the four-form 
Lagrange multiplier in order to get $\epsilon(y)$ type coefficient dynamically. 
We also introduce a compensator and a physical hypermultiplet whose 
scalar component is represented by the quaternion ${\cal A}$ and $\Phi$ 
respectively. These are gauged by the vector multiplets with the charge 
assignment of $\big(\,T_Z,T_X,T_S\,\big){\cal A}=
\big( 0,\tilde{q},\textstyle{-\frac{3}{2}}k \big)i\sigma_3{\cal A}$ 
and $\big(\,T_Z,T_X,T_S\,\big)\Phi= \big( 0,q,c \big)i\sigma_3\Phi$. 
We set $\tilde{q}=0$ in order not the ${\cal V}_X$ to take part in 
the $R$-gauging. We remark that the charge $k$ determines the AdS 
curvature of the background geometry and the $c$ gives bare kink mass 
for the hypermultiplet. In addition, we introduce the linear (four-form) 
multiplet $L|_{{\cal V}_Z}$ as a Lagrange multiplier constraining ${\cal V}_S$, 
whose vector and real scalar component is dualized into the three- and 
four-form filed respectively under the background of ${\cal V}_Z$. 

We start from the off-shell SUGRA Lagrangian with Weyl multiplet, 
the above three vector multiplets and two hypermultiplets, plus 
the Lagrange multiplier terms with the four-form multiplet which 
was derived in Ref.~\cite{Fujita:2001bd}. The norm function 
of our model is defined as 
$${\cal N}=\alpha^3 -{\textstyle \frac{1}{2}} \alpha \beta^2 
+{\textstyle \frac{1}{2}} \xi_{FI} \alpha \beta \gamma,$$ 
which after integrating out four-form field results in 
${\cal N}= \alpha^3 -{\textstyle \frac{1}{2}} \alpha \beta^2 
+{\textstyle \frac{1}{2}}\xi_{FI} \epsilon(y) \alpha^2 \beta$. 
The action after integrating out four-form multiplet 
is given by the sum of bulk and boundary contributions. 
Besides the usual bulk action, the brane action (bosonic part) 
is induced as 
\begin{eqnarray}
e_{(4)}^{-1} {\cal L}_{\rm brane} 
&=& \Big[ {\textstyle \frac{1}{2}}\xi_{FI}\alpha^2 
({\rm tr}[i\sigma_3Y^X]+\partial_4 \beta) 
\nonumber \\ &&
-2\alpha(3k+{\textstyle \frac{3}{2}}{\rm tr}[\Phi^\dagger \Phi] 
+c{\rm tr}[\Phi^\dagger \sigma_3 \Phi \sigma_3]) \Big] 
\big( \delta(y)-\delta(y-\pi R) \big), 
\nonumber
\end{eqnarray}
where we find the boundary FI-term for the ${\cal V}_X$ multiplet 
in the first line with the coefficient $\xi_{FI}$ defined in the 
norm function. (The full result is shown in Ref.~\cite{Abe:2004yk}.) 
We note that this brane action proportional to 
$\delta(y)-\delta(y-\pi R)=\partial_y \epsilon(y)/2$ 
is the consequence of the four-form mechanism~\cite{Bergshoeff:2000zn}. 

On the 4D Poincar\'e invariant background geometry 
$ds^2=e^{2K}\eta_{\underline\mu \underline\nu} 
dx^{\underline\mu} dx^{\underline\nu} -dy^2$, 
the 4D energy density is calculated as 
\begin{eqnarray}
E=\int \,dy \,\, e^{4K}\left(
\frac{1}{2} a_{IJ}D^ID^J 
+\frac{2}{1+v^2}|F|^2 -6|\kappa|^2 \right), 
\nonumber
\end{eqnarray}
where $I,J=(Z,X)$, 
$a_{IJ}=-\frac{1}{2}
\frac{\partial^2 \ln {\cal N}}{\partial M^I \partial M^J}$ 
and we assume $\langle \Phi^{(0)} \rangle=v$ and 
$\langle \Phi^{(r)} \rangle=0$ for the quaternionic hyperscalar field 
$\Phi=\Phi^{(0)} \mathbf{1}_2+i\sum_{r=1}^{3} \Phi^{(r)}\sigma_r$. 
The superconformal gauge fixing ${\cal N}=1$ reduces the number 
of scalar field as 
\begin{eqnarray}
\alpha(\phi) &=& 
(1+\xi_{FI}^2\epsilon^2(y)/8)^{-1/3} 
\cosh^{2/3} (\phi), 
\nonumber \\
\beta(\phi) &=&  \alpha(\phi) 
\big[ (2+{\textstyle \frac{1}{4}}\xi_{FI}^2 
\epsilon^2(y))^{1/2} \tanh (\phi) 
+{\textstyle \frac{1}{2}} \xi_{FI} \epsilon(y) \big], 
\nonumber
\end{eqnarray}
where we find that $\beta$ almost carries the physical 
degree of freedom $\phi$ at the leading order. 
The $D$, $F$ and $\kappa$ are the gaugino, hyperino and gravitino 
($N=1$) Killing order parameters respectively, 
\begin{eqnarray}
\kappa &=& \partial_y K - {\cal P}/3, \qquad 
F \ = \ \partial_y v - (q \beta + c\epsilon(y) \alpha-{\cal P}/2)v, 
\nonumber \\
D^I &=& {\cal N}^{IJ} \Big( 
\partial_y {\cal N}_J +2{\cal N}_J{\cal P}/3 
+6k\epsilon(y)\delta_J^{\ Z} 
+4 \big( (3k/2+c)\epsilon(y) \delta_J^{\ Z}
+q \delta_J^{\ X} \big) v^2 \Big), 
\nonumber
\end{eqnarray}
where 
${\cal P}=-2 \left[ {\textstyle \frac{3}{2}}k\epsilon(y)\alpha 
+\left( \left( {\textstyle \frac{3}{2}}k+c \right) \epsilon(y)\alpha 
+q\beta \right)v^2 \right]$ 
and ${\cal N}_{IJ}{\cal N}^{JK}=\delta_I^{\ K}$. 
We have a ($\beta$-independent) boundary FI-contribution 
in the first term ${\cal N}^{XX} \partial_y {\cal N}_X$ in $D^X$. 
We should note that the stationary condition of the energy functional 
is just given by Killing conditions $\kappa=F=D^I=0$.

\section{Some physical consequences}
Above we succeeded to formulate boundary FI-term off-shell. 
In this section we show some physical consequences of the FI-term. 
Combining $\kappa=0$ and $D^I=0$, we find 
\begin{eqnarray}
e^{-2K} \partial_y (e^{2K}{\cal N}_I) 
&=& -6k\epsilon(y) \delta_I^{\ Z}
-4((3k/2+c)\epsilon(y)\delta_I^{\ Z}+q\delta_I^{\ X})v^2. 
\label{eq:killing}
\end{eqnarray}
Obviously $F=0$ is satisfied by $v=0$, and for such hyperscalar 
vacuum value, Eq.~(\ref{eq:killing}) for $I=X$ is satisfied by 
${\cal N}_X=-\alpha \beta+\xi_{FI}\epsilon(y)\alpha^2/2=0$ 
resulting 
\begin{eqnarray}
\beta(\phi) &=& \frac{1}{2}\xi_{FI}\epsilon(y)\alpha(\phi). 
\label{eq:alphabeta}
\end{eqnarray}
With this relation, ${\cal N}=1$ condition gives 
$\alpha(\phi)=(1+\xi_{FI}^2 \epsilon^2(y)/8)^{-1/3}$ that means 
\begin{eqnarray}
\langle \phi \rangle &=& 0. 
\nonumber
\end{eqnarray}
With these VEVs of $\phi$ and $v$, the $I=Z$ component of 
Eq.~(\ref{eq:killing}) yields 
$$K \simeq -k|y|,$$ 
and we find from Eq.~(\ref{eq:alphabeta}) that 
\begin{eqnarray}
\langle \beta \rangle 
&=& \frac{1}{2}\xi_{FI}\epsilon(y)\alpha(\phi=0)
\ = \ \frac{1}{2}\xi_{FI}\epsilon(y)
+{\cal O}(\xi_{FI}^2). 
\label{eq:betavev}
\end{eqnarray}
From this result, we conclude that there exists a supersymmetric 
vacuum with the FI-term $\xi_{FI} \ne 0$, and at ${\cal O}(\xi_{FI})$ 
the only vacuum deformation induced by the FI-term is the nonvanishing 
VEV of $\beta$ given by Eq.~(\ref{eq:betavev}) irrespective of the 
value of $k$, i.e., {\it irrespective of whether the background 
geometry is flat or AdS}~\cite{Correia:2004pz}. 

\section{Summary}
We have shown an exact formulation of the boundary FI-term 
in the fully local context. We utilized off-shell formulation with 
four-form Lagrange multiplier to realize the bulk odd operator 
which is relevant to the boundary FI-term in the orbifold SUGRA. 
We analyzed the BPS stationary conditions of the scalar potential 
with the FI-term $\xi_{FI} \ne 0$, and found that there is a 
supersymmetric vacuum. At ${\cal O}(\xi_{FI})$, the only vacuum 
deformation induced by the FI-term is the nonvanishing VEV of 
$\beta$ which generates the kink mass for the charged hypermultiplets. 
This result is irrespective of whether the background geometry 
is flat or AdS~\cite{Correia:2004pz}.

\section{Acknowledgements}
The author would like to thank K.~Choi and I.~W.~Kim for 
the collaboration~\cite{Abe:2004yk} which forms the basis 
of this talk. 

\bibliographystyle{plain}

\end{document}